# Influence of Centre Body on the Dynamics of Isothermal Flow Swirl Combustor

Ratnesh Pathak[1], Nitesh Kumar Sahu[1*], and Pradeep Kumar Sonkar[1]

[1]Department of Fuel, Minerals and Metallurgical Engineering, Indian Institute of Technology (ISM) Dhanbad, Jharkhand -826004, India

24mt0354@iitism.ac.in; *nitesh@iitism.ac.in; 24mt0300@iitism.ac.in

## ABSTRACT

This study examines the effect of centre-body geometry on the dynamics of an isothermal, non-reacting swirl combustor through computational fluid dynamics (CFD) simulations. Two different central body shapes were considered in a lab-scale combustor configuration, modelled as transient, incompressible flow using the SST k–ω turbulence model. The numerical model was validated against experimental velocity data from literature to ensure accuracy. Cross-spectral analysis techniques were employed to characterise the coherent dynamics of the flow, providing insight into the influence of geometry on unsteady swirl dynamics

## 1. INTRODUCTION

Swirl-stabilized combustion is a common strategy in propulsion and power systems, valued for its ability to promote rapid mixing and maintain stable flames under lean operating conditions. When the applied swirl exceeds a certain critical level, the flow undergoes vortex breakdown [4,5], generating a central recirculation zone (CRZ) [6,7]. This zone recirculates hot gases and reactive species toward the flame root, improving ignition stability and extending lean blow-off limits. The breakdown appears as bubble-type or spiral-type [4,5] flow structure, accompanied by a precessing vortex core (PVC) [1,4] that induces large-scale unsteadiness. Such coherent motions strongly influence flame shape, and pollutant formation. A clear understanding of these coupled flow–flame dynamics is therefore essential for designing swirl combustors that achieve low emissions while maintaining robust performance across a wide operating range. However present study limited to isothermal, non-reacting flow.

## 2. LITERATURE REVIEW AND OBJECTIVE

Swirling flows in combustors promote flame stability by creating a CRZ through vortex breakdown. The swirl number $S$ in Eq. (1) is a key parameter controlling this process. Manoharan *et al.* [1] examined turbulent swirling jet where they primary deal with interaction of vortex core deformation with the precession of vortex breakdown. Their results showed that increasing swirl significantly altered the flow field, enlarging the recirculation zone, strengthening shear layers, and generating a precessing vortex core. Moreover, there are many other associative studies cannot find any other investigation of the impact of centre body shape of swirler on the unsteady flow characteristics in isothermal swirl combustor. Accordingly presenting the same in our present work.

$$S = \frac{\int_0^R \rho\, u_a u_\theta\, r^2\, dr}{R \int_0^R \rho\, u_a^2\, r dr} \qquad (1)$$

where $S$ is swirl number, $R$ is the inlet or reference radius, $\rho$ is fluid density, $u_a$ is axial velocity component, $u_\theta$ is azimuthal velocity component and $r$ is radial coordinate .

## 3. METHODOLOGY

### 3.1 Geometry and Computational Modelling

The computational setup follows the geometry of a lab-scale swirl combustor configuration reported by Taamallah *et al.* [2]. It features a straight inlet pipe that transitions abruptly into a cylindrical combustion chamber. The flow enters axially through a swirler equipped with eight blades set at a 45° angle, mounted on a central body. Two centre-body geometries were examined: a bullet-shaped profile and a cylindrical profile., which generates the swirl. The inlet section measures 0.038 m in diameter and 0.045 m in length, while the swirler section has a length of 0.030 m. The combustion chamber begins with a diameter of 0.038 m and length of 0.045 m, followed by an expanded section of 0.076 m diameter and 0.225 m length. Reynolds number of approximately 20,000. The flow is simulated under transient, incompressible, and isothermal conditions using the Reynolds-Averaged Navier–Stokes (RANS) framework. Turbulence is captured with the shear-stress-transport (SST) k–ω model, and all simulations are performed with ANSYS Fluent 2025R1 Commercial software.

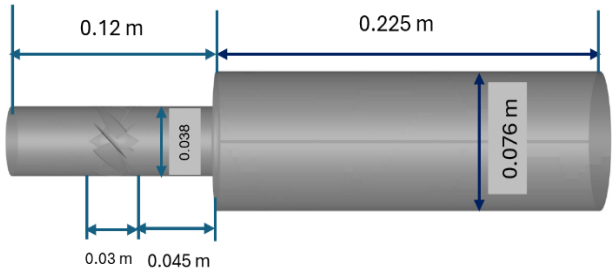

**Figure 1: Schematic of the combustor geometry (Taamallah *et al*. [4]).**

### 3.2 Boundary Condition

A velocity inlet boundary condition was applied at the inlet, specifying an inflow velocity of 8.2 m/s. while the outlet was assigned an ambient pressure outlet boundary condition. All walls were assigned no-slip boundary condition.

### 3.3 Numerical Method and Turbulence Modelling

The unsteady flow field was computed by solving the Reynolds-Averaged Navier–Stokes (RANS) equations with the shear-stress-transport (SST) *k–ω* turbulence model, chosen for its capability to capture strong adverse pressure gradients and flow separation. Simulations were performed in ANSYS Fluent 2025R1 using the pressure-based coupled solver, with second-order upwind discretization for convective terms and the PRESTO! scheme for pressure interpolation two faces. The transient simulation was first advanced over Three flow-through times to allow the decay of initial transients, using a time-step size of 0.001 s and Forty iterations per time step, Convergence was achieved when scaled residuals dropped below $1\times10^{-4}$. The computations were then continued for additional Ten flow-through times with the same time-step size, and for generating time-averaged results, final 3 flow steps data were extracted for temporal analysis.

### 3.4 Grid Independence Study

A grid independence study was conducted by comparing the axial distribution of turbulent kinetic energy (*k*) across three successively refined meshes. All computational meshes consisted entirely of structured hexahedral elements, as shown in Figures 2(a) and 2(b). For the bullet-shaped bluff body configuration as well as the cylindrical shape bluff body configuration, a mesh with approximately 0.5 million hexahedral elements was selected, as further refinement produced negligible differences in the turbulent kinetic energy (TKE) profiles. The chosen mesh maintained an aspect ratio of 20 and an orthogonal quality of 0.12, providing a suitable compromise between numerical accuracy and computational cost. In both cases, the near-wall resolution corresponded to a wall $y^+$ of approximately 30, ensuring adequate boundary-layer resolution and accurate representation of near-wall flow behaviour.

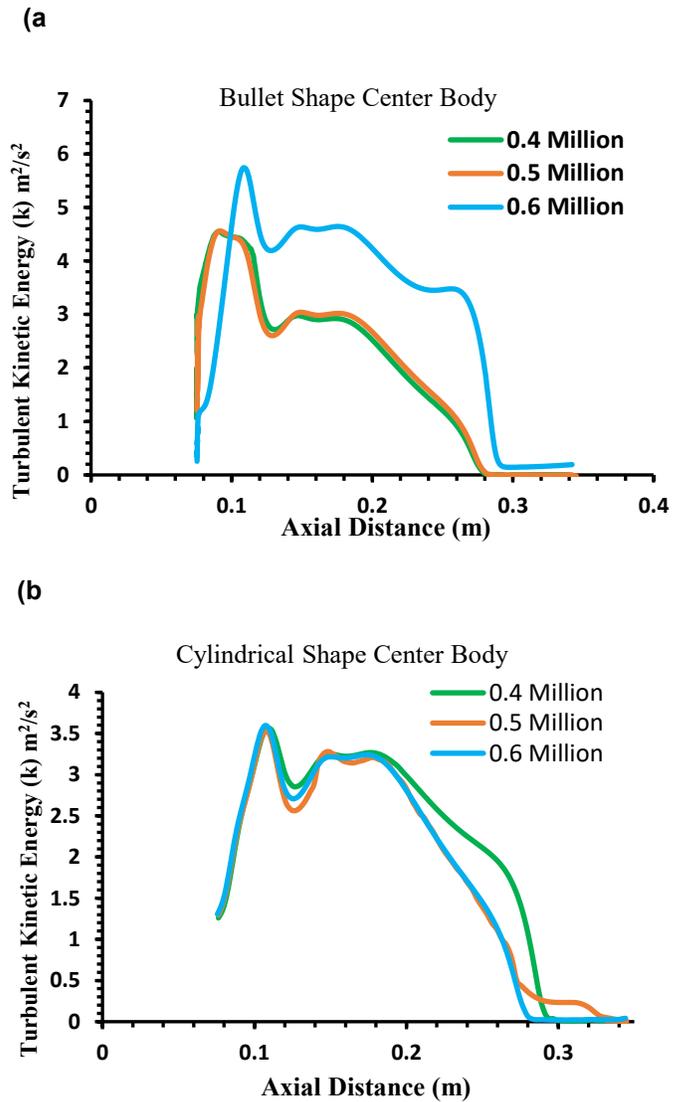

**Figure 2: - Gride Independence Study (a) Bullet shape centre body (b) cylindrical shape centre body**

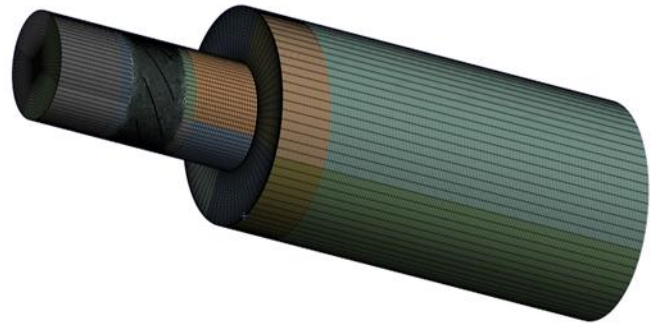

**Figure 3: - Computational Mesh with 0.5 million Hexahedral Elements**

### 3.5 Validation study

To evaluate the accuracy of the numerical model, the simulated radial distributions of axial velocity were compared with the experimental measurements of Taamallah *et al.* [2]. The comparison, presented in Figures 4(a) and 4(b), corresponds to two downstream locations, $x/R = 0$ and 0.5, where $x$ is the axial distance from the sudden expansion and $R$ represents the combustor radius (0.038 m). The predicted velocity profiles show strong consistency with the experimental trends, successfully capturing both the peak magnitudes and the overall shape of the distribution. This level of consistency demonstrates that the numerical setup can reliably reproduce the key flow characteristics, providing confidence in extending the simulations for further parametric studies and detailed flow investigations.

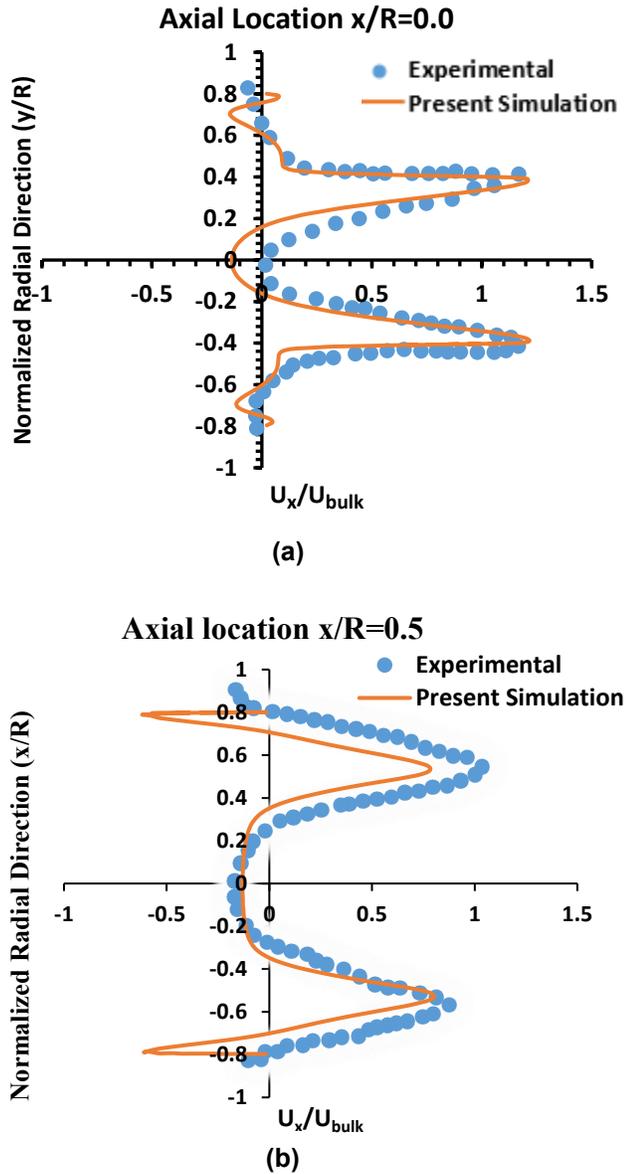

**Figure 4:** - Comparison of radial profiles of axial velocity with experimental data [2] at (a) x/R=0.0 (b) x/R= 0.5

### 4. RESULTS AND DISCUSSION

In this section, we examine the coherent dynamics of the vortex core in the investigated systems. The coherent dynamics of vortex core commonly precession imparts footprints on the axial component of turbulent kinetic energy. Maxima appears in the corresponding field usually recognized as the footprints of this precession. However, the axial component of turbulent kinetic energy can't be filtered out in our simulation with eddy-viscosity turbulence model so, we look for the maxima in contours of turbulent kinetic energy, shown in fig. 5. This simplification is suitable for our axial-flow-dominated system, as regions of maximum turbulent kinetic energy typically overlap with regions of maximum axial component of turbulent kinetic energy. Two diametrically opposite points in the figure below appear to possess maximum value in each of the investigated case, which are specified as follows.

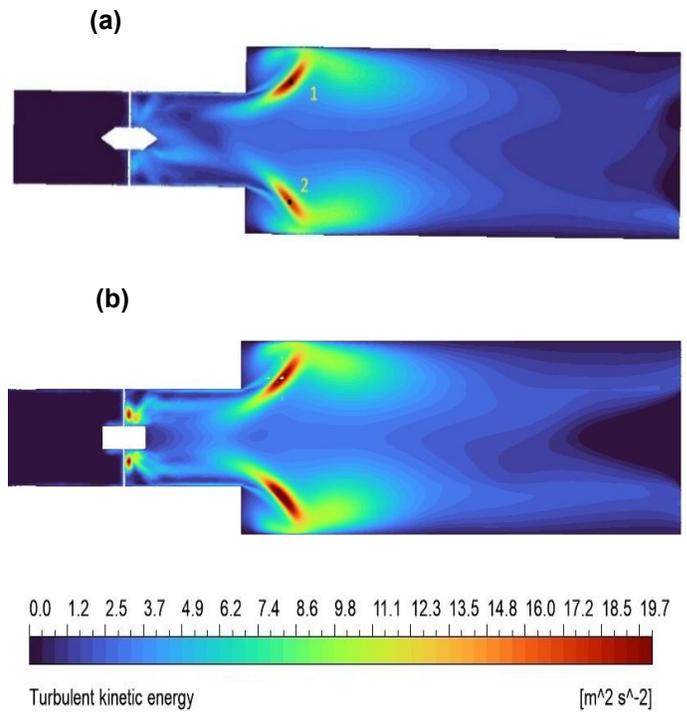

**Figure 5:** - Contours of Turbulent kinetic energy (a) Bullet Shape Bluff body (b) Cylindrical shape Bluff body.

In Figures 5(a) and 5(b), two monitoring points are indicated for each configuration. For the bullet-shaped central body, the points are located at $(0, 0.143, −0.0235)$ and $(0, 0.143, 0.0235)$, while for the cylindrical central body they are at $(0, 0.141, 0.0233)$ and $(0, 0.141, −0.0233)$, where coordinates are given as (x, y, z). Axial velocity fluctuations at these points are recorded over three flow-through times. Cross-spectral analysis is then carried out for each case, following the methodology described by Manoharan et al. For this purpose, five ensembles of 120 points each are used by us, giving a total of 360-time steps covering the three flow-through times

The cross-spectral analysis [1,3] yields two outputs: the amplitude spectrum, showing the strength of oscillations at different frequencies, and the corresponding phase difference between the two monitoring points. As illustrated in Figures 6(a) and 6(b), the amplitude peaks highlight the dominant frequencies present in the flow, while the phase plots indicate the relative timing between signals. In swirling flows, phase differences near $\pi$, $3\pi$, or $5\pi$ (and their negative counterparts) typically correspond to odd azimuthal modes, with the mode number estimated from $\Delta\theta/\pi$ Similarly, phase differences around $2\pi$, $4\pi$, etc., indicate even modes.

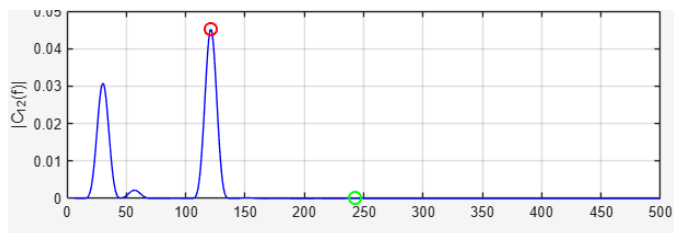

(1)

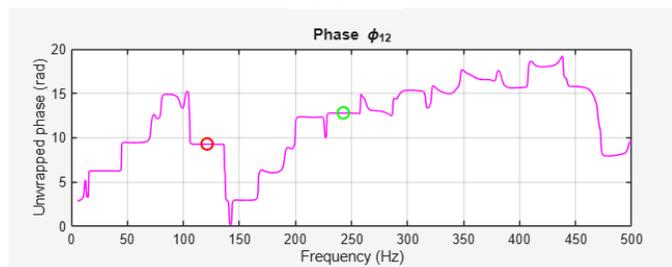

(2)

**Figure 6(a): - (1) Cross-spectral magnitude (2) unwrapped phase between two monitoring points for bullet-shaped Central Body.**

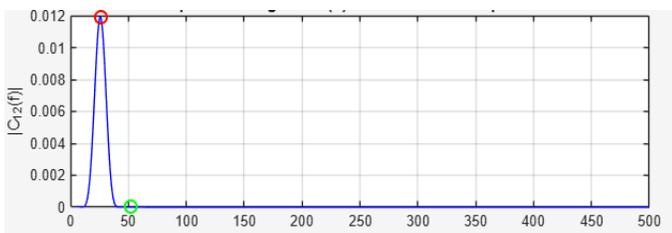

(1)

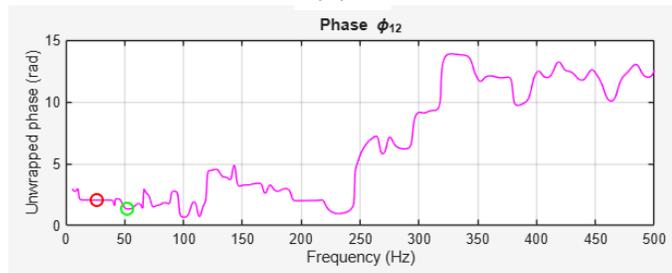

(2)

**Figure 6(b): - (1) Cross-spectral magnitude (2) unwrapped phase between two monitoring points for Cylindrical shaped Central Body.**

The results indicate that, for the bullet-shaped bluff body configuration (Figure 6a), the maximum cross-spectral amplitude occurs at a frequency of 121 Hz. The corresponding phase difference, marked in red at this frequency, yields a mode number of m=−3. In contrast, for the cylindrical bluff body configuration (Figure 6b), the highest amplitude is observed at 25 Hz, with the red-marked phase indicating a mode number of m=−1. The mode numbers are determined from the phase difference at the frequency of maximum amplitude, following the relation m≈$\Delta\theta/\pi$.

This result shows a significant change in the dominant frequency corresponding to the maximum amplitude. In the first case, if we neglect the small structural changes, the mode number can be considered as m=−1. From this, we conclude that for m=−1, there is a large difference in frequency between the two configurations, with the cylindrical shape suppressing the frequency of the precessing vortex core motion. It should be noted that the exact reason for this suppression is not yet understood.

## 5. CONCLUSIONS

This study numerically investigates the effect of centre-body shape on the unsteady characteristics of an isothermal swirl combustor flow. Analysis of turbulent kinetic energy and cross-spectral analysis reveals that the frequency of precession of vortex core in bullet-shaped centre-body configuration is higher compared to the corresponding frequency in the cylindrical centre-body configuration. Thus, it is evident the cylindrical shaped centre-body in a swirl combustor suppresses the precessing dynamics of vortex core compared to the bullet shaped centre-body.

## ACKNOWLEDGEMENTS

The present work is supported by the Faculty Research Scheme (FRS) Grant; FRS (217)/2024-25/FMME of IIT-ISM Dhanbad. We further thank TEXMiN Hub for their technical support.